\documentclass[conference]{IEEEtran}
\IEEEoverridecommandlockouts
\usepackage{cite}
\usepackage{amsmath,amssymb,amsfonts}
\usepackage{algorithmic}
\usepackage{graphicx}
\usepackage{textcomp}
\usepackage{xcolor}
\usepackage{hyperref}

\usepackage{booktabs}
\usepackage{multirow}
\usepackage{enumitem}

\def\BibTeX{{\rm B\kern-.05em{\sc i\kern-.025em b}\kern-.08em
    T\kern-.1667em\lower.7ex\hbox{E}\kern-.125emX}}

\begin{document}

% \title{Conference Paper Title*\\
% {\footnotesize \textsuperscript{*}Note: Sub-titles are not captured for https://ieeexplore.ieee.org  and
\title{Stimulus Modality Matters: Impact of Perceptual Evaluations from Different Modalities on Speech Emotion Recognition System Performance}

\author{\IEEEauthorblockN{Huang-Cheng Chou}
\IEEEauthorblockA{\textit{Department of Electrical Engineering} \\
\textit{National Tsing Hua University}\\
Hsinchu City, Taiwan \\
huangchengchou@gmail.com}
\and
\IEEEauthorblockN{Haibin Wu, Hung-yi Lee}
\IEEEauthorblockA{\textit{Department of Electrical Engineering} \\
\textit{National Taiwan University}\\
Taipei City, Taiwan \\
\{f07921092, hungyilee\}@ntu.edu.tw}
\and
\IEEEauthorblockN{Chi-Chun Lee}
\IEEEauthorblockA{\textit{Department of Electrical Engineering} \\
\textit{National Tsing Hua University}\\
Hsinchu City, Taiwan \\
cclee@ee.nthu.edu.tw}
% \and
% \and
% \and
}

\maketitle

\begin{abstract}
\small
    Speech Emotion Recognition (SER) systems rely on speech input and emotional labels annotated by humans. However, various emotion databases collect perceptional evaluations in different ways. For instance, the IEMOCAP dataset uses video clips with sounds for annotators to provide their emotional perceptions. However, the most significant English emotion dataset, the MSP-PODCAST, only provides speech for raters to choose the emotional ratings. Nevertheless, using speech as input is the standard approach to training SER systems. Therefore, the open question is the emotional labels elicited by which scenarios are the most effective for training SER systems. We comprehensively compare the effectiveness of SER systems trained with labels elicited by different modality stimuli and evaluate the SER systems on various testing conditions. Also, we introduce an all-inclusive label that combines all labels elicited by various modalities. We show that using labels elicited by voice-only stimuli for training yields better performance on the test set, whereas labels elicited by voice-only stimuli.
\end{abstract}

\begin{IEEEkeywords}
speech emotion recognition, the effects of stimulus modality, the ambiguity of emotions
\end{IEEEkeywords}

\small

\section{Introduction}
\label{s:intro}

%Prior studies investigated Speech Emotion Recognition (SER) systems using different types of stimuli. 
Prior studies trained Speech Emotion Recognition (SER) systems using labels elicited by the different types of stimuli. There are two main ways to obtain labels. One is giving raters audio-only emotional stimuli and letting them assign labels. For instance, the MSP-PODCAST emotion dataset \cite{Lotfian_2019_3} uses this scenario. The other is giving raters audio-visual emotional stimuli and letting them provide labels, and the IEMOCAP \cite{Busso_2008_5} emotion dataset is in the condition. While some papers have utilized labels derived from audio-visual stimuli \cite{Goncalves_2022} to train SER systems, others have focused on training SER models with labels elicited by audio-only stimuli \cite{Chou_2023}.
This variation in methodology raises an intriguing open question: \textbf{are the emotional labels elicited by multi-modal emotional stimuli different from those used in training SER systems with audio-only inputs?}

%\textbf{is the emotional labels elicited by multi-modal emotional stimuli different from training SER systems than those elicited by audio-only?} 

The emergence of speech self-supervised learning models (SSLMs) has significantly propelled advancements across a wide array of speech-related tasks \cite{Yang_2021}, including SER. The current state-of-the-art frameworks in SER are primarily built upon these SSLMs \cite{Wagner_2023}. To thoroughly investigate the research question concerning the advantage of utilizing labels elicited by multi-modal or single-modal emotional stimuli for training SER models, we conducted extensive experiments using SSLMs. In our approach, we trained SER systems using labels derived from various modalities, including audio-only, facial-only, and audio-visual inputs. This training utilized the S3PRL toolkit \cite{Yang_2021}, which encompasses 14 self-supervised learning models (SSLMs). Our findings highlight significant differences in the performance of SER systems trained with labels elicited from these modalities when tested under three distinct conditions: \textbf{audio-only, facial-only, and audio-visual label elicitation}.

%Our approach involved training SER systems using labels elicited from various modalities, such as audio-only, facial-only, and audio-visual. This training was facilitated by leveraging the S3PRL toolkit \cite{Yang_2021}, encompassing 14 SSLMs. We revealed notable differences in the performances of SER systems trained with the labels elicited by the mentioned modalities across three testing conditions defined by labels elicited from \textbf{audio-only, facial-only, and audio-visual modalities}. 

We initiated a cross-testing experiment to discern the most productive approach to training SER systems. 
For different labeling processes using various stimuli, we use one type of label for training and all label types for testing.
For instance, we trained SER systems using labels elicited by audio-only stimuli. Then, we evaluated these models using test sets labeled with audio-only, facial-only, and audio-visual stimuli. Furthermore, we introduced an innovative \emph{all-inclusive} label set that combines labels elicited by audio-only, facial-only, and audio-visual stimuli to train the SER systems. The SER systems trained with this \emph{all-inclusive} label set outperformed those trained with labels elicited by uni-modal or multi-modal emotional stimuli on the facial-only and audio-visual conditions. The SER systems trained with the voice-only label set achieved the best performance on the voice-only testing condition.

In conclusion, our work makes three contributions as follows. The source code is available\footnote{https://github.com/EMOsuperb/Stimulus-Modality-Matters}.

\begin{itemize}[noitemsep,topsep=1pt]
    \item We presented an exhaustive comparative analysis of SER systems trained with labels elicited by various annotation conditions (e.g., audio-visual stimuli) and evaluated on test sets with labels elicited by different modalities (e.g., voice-only stimuli). This analysis provides valuable insights into the performance of SER systems under different training and testing annotation conditions.
    \item We introduce a novel \emph{all-inclusive} label for training SER systems. Our results demonstrate that this label set shows promise for improving the performance of SER systems on the test set whose labels elicited by face-only and audio-visual modalities scenarios. 
    This finding highlights the potential of leveraging information from multiple annotation conditions to enhance the accuracy of SER systems.
    \item We are the first to reveal that training SER systems using labels elicited by audio-only stimuli is better than using labels elicited by audio-visual stimuli based on our extensive experimental results. Our findings indicate that focusing on audio cues alone during labeling is more effective for training SER in audio-only contexts, and the findings draw a connection to the fact that recent benchmark databases (such as MSP-PODCAST) use audio-only stimuli for labels. 
\end{itemize}

\section{Background and Related Work}
\label{s:background}

Emotion perception is inherently multifaceted, influenced by various sensory inputs such as auditory signals, facial expressions, and a combination of audio-visual cues \cite{Paulmann_2011}. Consequently, the field of emotion recognition has evolved to include a diverse array of systems, each focusing on different modalities: facial emotion recognition \cite{Kim_2021}, text emotion recognition \cite{Kumar_2022}, speech emotion recognition \cite{Riera_2019,Abdelwahab_2018}, and multi-modal (e.g., audio-visual \cite{Ma_2021,Lei_2023} or speech and text \cite{Lin_2023_3}) emotion recognition. \textbf{This study primarily seeks to advance SER systems that rely solely on speech as the input modality}. 

Much research has traditionally favored training SER models using labels derived from multi-modal stimuli, particularly audio-visual inputs. The IEMOCAP corpus \cite{Busso_2008_5}, one of the most influential datasets in SER research, exemplifies this approach by collecting labels through audio and visual stimuli. Recent trends, however, have shown an increasing shift toward using audio-only stimuli for collecting emotional labels in SER tasks. The MSP-PODCAST database \cite{Lotfian_2019_3} is the most extensive annotated emotional corpus in English and relies exclusively on audio stimuli for label collection. This shift marks an emerging need to evaluate the efficacy of labels derived from varying stimuli types for training SER models. Investigating this research question could provide valuable insights into optimizing SER models for more accurate and efficient emotion detection.

\section{Methodology}
\label{s:method}

% P1: Motivation and existing resources
Most prior SER researches have mainly relied on annotations elicited by audio-visual stimulus as their learning objective \cite{Goncalves_2022}, or the prior study did not specify labels elicited by which modalities they used \cite{Chen_2024}.
%However, the labels elicited by which modalities can lead to better performances have yet to be explored. 
However, the relationship between the modalities from which labels are elicited and the resulting performance improvements has not yet been thoroughly investigated.
To answer the question, we aim to compare the performances of SER systems trained with labels elicited by different modalities (e.g., face-only or audio-only) across various testing conditions according to modality stimulus.

\begin{figure}[!b]
\centering
\vspace{-5mm}
\includegraphics[width=0.7\linewidth]{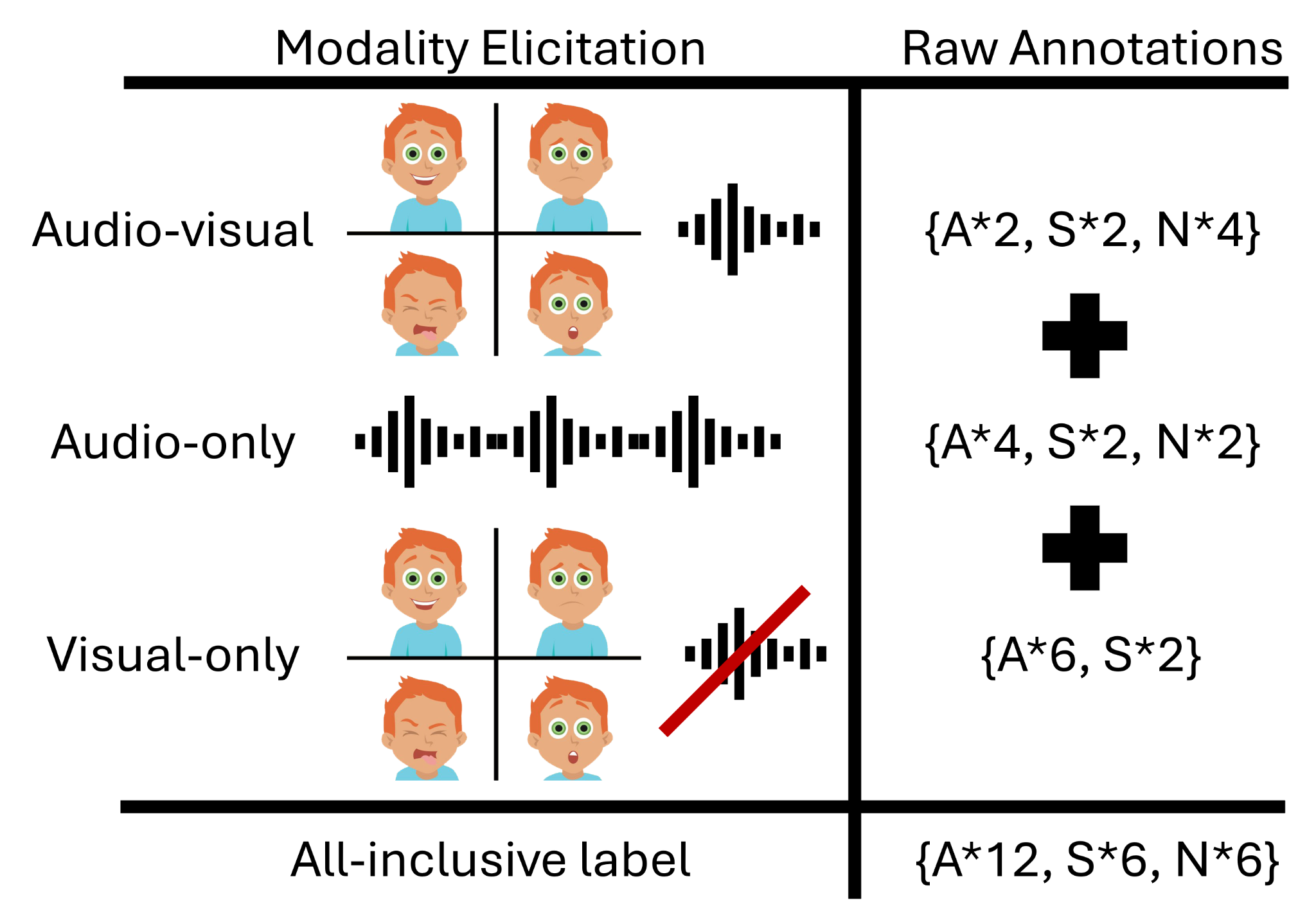}
\vspace{-3mm}
\caption{The figure shows the multi-modal emotional stimulus in the CREMA-D emotion corpus and the proposed all-inclusive label.} 
\label{fig:showcase}
\end{figure}

\subsection{Labels Elicited by Multi-modal Emotional Stimulus}

The annotation process in the CREMA-D corpus \cite{Cao_2014} introduced in Section~\ref{ss:dataset} encompassed three scenarios: \textbf{voice-only, face-only, and audio-visual settings}. 
In the voice-only scenario, annotators were presented solely with the audio component of the clips for their reference to label the data. 
Conversely, annotators were limited to observing actors' facial expressions without accompanying audio when assigning labels in the face-only setting.
On the other hand, the audio-visual setting provided annotators with a complete experience, enabling them to see the actors' faces and hear their voices.

\subsection{Proposed All-inclusive Label Set and Rationales}

% P2: Proposed method
% We propose an innovative approach for training SER systems. 
% This approach involves creating an \emph{all-inclusive} label set (\textbf{All}) that integrates labels elicited from uni-modal and multi-modal emotional stimuli. 

% Fig. \ref{fig:showcase} illustrates our proposed innovative approach for training SER systems, involving creating an \emph{all-inclusive} label set (\textbf{All}) that integrates labels elicited from uni-modal and multi-modal emotional stimuli. 

Fig. \ref{fig:showcase} illustrates our proposed innovative approach for training SER systems, which involves creating an \emph{all-inclusive} label set (\textbf{All}) that integrates labels derived from uni-modal and multi-modal emotional stimuli.
By leveraging the comprehensive spectrum of emotional cues available from audio, visual, and audio-visual sources, we aim to harness the synergistic effect of multi-modal input to improve the performance of SER systems. This strategy is directly inspired by the inherent human capability to more accurately perceive and interpret emotions when multiple modal cues are available.
%This strategy is directly motivated by the natural human ability to perceive better and interpret emotions when more modal cues are available.

% How to create the labels and differences in the training and testing stages
In the example of Table~\ref{tab:comparisions}, we summarize how to convert the raw annotations into the training/testing labels generated by the different modalities. 
The \emph{all-inclusive label} (\textbf{All}) considers all labels elicited by voice-only, facial-only, and audio-visual stimuli. 
We consider a six-class emotion task, including anger (A), disgust (D), fear (F), happiness (H), sadness (S), and a neutral state (N). For voice-only elicitation, the annotations are \{A*2, S*2, N*4\}; for facial-only elicitation, the annotations are \{A*4, S*2, N*2\}; and for audio-visual elicitation, the annotations are \{A*6, S*2\}. The proposed \emph{all-inclusive label} integrates all ratings, resulting in annotations of \{A*12, S*6, N*6\}. 

% It is important to note that the labels are distributional labels for the training stage, and they are converted into binary vectors when the values are higher than the defined threshold described in Section~\ref{ss:metric}, which is 1/C, where C is the number of emotions. 
Importantly, the labels used during the training stage are distributional and are converted into binary vectors when their values surpass the defined threshold outlined in Section~\ref{ss:metric}, which is 1/C, where C is the number of emotions.
In this work, we have six emotions in total (threshold = 1/6), so the testing label of the audio-visual scenario (1,1,0,0,0,0) is different from others (1,1,0,0,0,1) after applying the threshold method introduced in \cite{Riera_2019}. We follow \cite{Chou_2022} to allow the samples to have more than one emotion to reflect the nature of emotion perception that could involve mixed emotions from the psychology perspective \cite{Cowen_2021}.

\begin{table}[!t]
\centering
\fontsize{7}{9}\selectfont
\caption{Overview of one sample in the CREMA-D. The \textbf{A}, \textbf{S}, and \textbf{N}, are anger, sadness, and neutral emotions, respectively. The number means the count of emotions. For instance, \textbf{A*2, S*2, and N*4} means A, A, S, S, N, N, N, N}
\vspace{-3mm}
\begin{tabular}{c|l|c}
\hline
\multirow{3}{*}{Voice-only}    & Raw Annotation & A*2, S*2, N*4               \\ \cmidrule{2-3}
                               & Label for Training Stage& (0.25,0.25,0.0,0.0,0.0,0.5) \\ 
                               & Label for Testing Stage& (1,1,0,0,0,1)               \\ \midrule
\multirow{3}{*}{Facial-only}   & Raw Annotation & A*4, S*2, N*2               \\ \cmidrule{2-3}
                               & Label for Training Stage & (0.5,0.25,0.0,0.0,0.0,0.25) \\
                               & Label for Testing Stage  & (1,1,0,0,0,1)               \\ \midrule
\multirow{3}{*}{Audio-Visual}  & Raw Annotation & A*6, S*2                    \\ \cmidrule{2-3}
                               & Label for Training Stage & (0.75,0.25,0.0,0.0,0.0,0.0) \\
                               & Label for Testing Stage  & (1,1,0,0,0,0)               \\ \midrule
\multirow{3}{*}{All-inclusive} & Raw Annotation & A*12, S*6, N*6              \\ \cmidrule{2-3}
                               & Label for Training Stage & (0.5,0.25,0.0,0.0,0.0,0.25) \\
                               & Label for Testing Stage  & (1,1,0,0,0,1)               \\ \bottomrule
\end{tabular}
\vspace{-5mm}
\label{tab:comparisions}
\end{table}

\subsection{SSLMs-based SER Framework}

We adopt SER models using 14 SSLMs as the backbone models following the EMO-SUPERB settings\footnote{https://github.com/EMOsuperb/EMO-SUPERB-submission} \cite{Wu_2024} to train SER systems. We use SSLMs as they achieve SOTA results in SER.
We leverage two mainstream categories of SSLMs, pre-trained using generative loss, \textbf{DeCoAR 2} \cite{Ling_2020}, Autoregressive Predictive Coding (\textbf{APC}) \cite{Chung_2019}, \textbf{VQ-APC} \cite{Chung_2020}, Non-autoregressive Predictive Coding (\textbf{NPC}) \cite{Liu_2020}, \textbf{TERA} \cite{Liu_2021}, and \textbf{Mockingjay} \cite{Liu_2020_2}), and discriminative loss (\textbf{XLS-R-1B}) \cite{Babu_2021}, \textbf{WavLM Large} \cite{Chen_2022}, \textbf{Data2Vec-A} \cite{Baevski_2022}, \textbf{Hubert Large} \cite{Hsu_2021}, \textbf{wav2vec 2.0 Large} (\textbf{W2V2}) \cite{Baevski_2020}, VQ wav2vec (\textbf{VQ-W2V}) \cite{Baevski_2019}, wav2vec (\textbf{W2V}) \cite{Schneider_2019}, wav2vec 2.0 Robustness (\textbf{W2V2 R}) \cite{Hsu_2021_2} and Contrastive Predictive Coding (CPC) (\textbf{M CPC})\cite{Oord_2018}). Additionally, we include the log mel filterbank (\textbf{FBANK}).

\section{Experimental Settings}
\label{s:experiment}

\subsection{The CREAMA-D}
\label{ss:dataset}
The CREMA-D dataset, as introduced by Cao et al. \cite{Cao_2014}, encompasses high-quality audio-visual clips featuring performances by 91 professional actors.
This rich dataset comprises 7,442 clips in English, and every clip received annotations from at least six raters, who were allowed to select only one of the six emotions, including anger, disgust, fear, happiness, sadness, and a neutral state. 
The annotation process of the database contains three scenarios: voice-only, face-only, and audio-visual settings, as shown in Fig.~\ref{fig:showcase}, mentioned in section~\ref{s:method}. The database does not provide a standard partition\footnote{https://emosuperb.github.io/standardization.html}, so we use the defined partition provided by EMO-SUPERB \cite{Wu_2024}. In total, there were five sessions, and we reported average results.

However, there is a lack of clarity in existing literature regarding the specific labels used to train SER models, as noted in several prior SER studies \cite{Chen_2024}. 
This highlights the need for further investigation into the optimal stimuli for training SER models, potentially unlocking new insights into more effective SER methodologies.

\subsection{Class-balanced Objective Function}
We follow the EMO-SUPERB \cite{Wu_2024} to employ the Class-Balanced Cross-Entropy Loss (BCE) strategy as a loss function, initially proposed by Cui et al. \cite{Cui_2019}. The BCE method incorporates a weighting factor into the loss function, designed to recalibrate the loss based on the inverse frequencies of each class within the training dataset. This approach ensures that each class is given appropriate consideration during training, regardless of its frequency, thereby mitigating the challenges posed by uneven annotation distributions and leading to more robust and equitable SER system performance.

\vspace{-1mm}
\subsection{Evaluation Metrics and Confidence Intervals}

\label{ss:metric}
In our evaluation framework, we follow the EMO-SUPERB \cite{Wu_2024} and recent SER challenge \cite{Goncalves_2024} to utilize the macro-F1 score and F1 score, metrics that simultaneously assess recall and precision rates to provide a balanced measure of our SER systems' performance \cite{opitz2019macro}. 
This evaluation method is executed using the Scikit-learn library \cite{Fabian_2011}. 
Our evaluation process adopts a threshold-based approach \cite{Riera_2019} for scenarios involving multi-label classifications to accurately identify the target classes from the ground truth data. 

Specifically, a prediction for a particular class is deemed correct if its proportional representation among all predictions exceeds the threshold of ($1/C$), where $C$ is the total number of emotional courses under consideration. 
This strategic choice of threshold ensures that predictions are classified based on a fair representation criterion, aligning with methodologies previously described in the literature \cite{Riera_2019,Chou_2024}. 
%Utilizing this approach allows for a nuanced and precise calculation of macro-F1 scores, effectively capturing the performance of our SER systems in recognizing a wide range of emotional states. 
Notice that we collect the predictions from each partition defined in the study \cite{Wu_2024} and then measure the performance in macro-F1 score with the average and lower and upper bound of the confidence interval (CI) between 2.75\% and 97.5\% using the toolkit \cite{ferrer_2024}. All results are single-run with a fixed random seed number.

\subsection{Models Training and Choice}

\label{ss:training}
We employ the AdamW optimizer \cite{Loshchilov_2019} with a learning rate of 0.0001. The batch size is set to 32, and the models are trained for 50 epochs. The best-performing models are selected based on the lowest loss value on the development set. All experiments use two Nvidia Tesla V100 GPUs with 32 GB of memory, requiring approximately 84 GPU hours. Our work is built upon the S3PRL \cite{Yang_2021}\footnote{https://github.com/s3prl/s3prl}, which is implemented using PyTorch \cite{Adam_2017} and HuggingFace library \cite{Thomas_2020}.

\section{Results and Analysis}
%Different stimuli modalities can influence emotion perception. 
The CREMA-D database provides annotations for various stimulus modalities, including face-only (\textbf{Face}), voice-only (\textbf{Voice}), and audio-visual (\textbf{AV}). We propose the \emph{all-inclusive} label set (\textbf{All}), a combination of all modalities. To investigate the impact of these modalities on emotion perception, we calculate the multi-label distribution labels for six emotions, as described in Section~\ref{s:method}, using the annotations elicited by different modalities to answer some research questions as below.

\begin{table}[!t]
\fontsize{7}{9}\selectfont
\centering
\caption{The table summarizes the concordance correlation coefficient (CCC) between labels of various modalities. The \textbf{Face}, \textbf{Voice}, \textbf{AV}, and \textbf{All} represent face-only, voice-only, audio-visual, and a combination of all modalities, respectively.}
\vspace{-3mm}
\begin{tabular}{@{\hspace{0.2cm}}c@{\hspace{0.2cm}}c@{\hspace{0.2cm}}c@{\hspace{0.2cm}}c@{\hspace{0.2cm}}c@{\hspace{0.2cm}}}
\toprule
\textbf{Stimulus Modality} & \textbf{Face} & \textbf{Voice} & \textbf{AV} & \textbf{All} \\ \midrule
\textbf{Face}                & 1.000         & 0.459          & 0.805                 & \textbf{0.875}            \\
\textbf{Voice}               & 0.459         & 1.000          & 0.573                 & \textbf{0.745}            \\
\textbf{AV}        & 0.805         & 0.573          & 1.000                 & \textbf{0.913}            \\
\textbf{All}             & \textbf{0.875}         & \textbf{0.745}          & \textbf{0.913}                 & 1.000            \\ \bottomrule
\end{tabular}
\label{tab:cremad_modality_correlation}
\vspace{-3mm}
\end{table}

\textbf{What is a correlation between labels elicited by various modalities?} We employ the concordance correlation coefficient (CCC) \cite{Lawrence_1989} to assess the correlation between the averaged labels under different conditions, as presented in Table \ref{tab:cremad_modality_correlation}. 
Interestingly, the CCC between \textbf{Voice} and \textbf{AV} modalities is only 0.573, while the CCC between \textbf{Face} and \textbf{AV} is considerably higher at 0.805. 
Furthermore, the CCC between \textbf{Voice} and \textbf{All} (0.745) is lower than the CCC between \textbf{Face} and \textbf{All} (0.875), as well as the CCC between \textbf{AV} and \textbf{All} (0.913). 
The \textbf{All} modality exhibits overall higher correlations with other modalities, providing an additional justification for our proposal of the \emph{all-inclusive} label set (\textbf{All}), which considers all labels from all modalities. 
This approach ensures a comprehensive consideration of the information available when analyzing emotion perception.

\begin{table}[!b]
\fontsize{7}{9}\selectfont
\centering
\vspace{-5mm}
\caption{The table summarizes performances of the various SSLMS-based SER systems trained with the labels elicited by different modalities. The \textbf{Face}, \textbf{Voice}, and \textbf{AV} represent face-only, voice-only, and audio-visual, respectively. We use bold to represent the best performance according to each upstream model. All values are in macro-F1 scores.}
\vspace{-3mm}
\begin{tabular}{@{}c|c|ccc@{}}
% \begin{tabular}{@{\hspace{0.1cm}}c|@{\hspace{0.1cm}}c|@{\hspace{0.1cm}}c@{\hspace{0.2cm}}c@{\hspace{0.2cm}}c@{\hspace{0.2cm}}}
\toprule
\textbf{Upstream}   & \textbf{\#Pars. (M)}  & \textbf{Voice}  & \textbf{Face} & \textbf{AV}        \\ \midrule
\textbf{WavLM Large}      & 317              & \textbf{0.7117}  & 0.6366  & 0.7076    \\
\textbf{XLS-R-1B}   & 965              & 0.6764  & 0.6251  & \textbf{0.6960}    \\
\textbf{Hubert Large}    & 317             & 0.6746  & 0.6148  & \textbf{0.6823}    \\
\textbf{W2V2 Large}       & 317             & \textbf{0.6687}  & 0.5957  & 0.6555    \\
\textbf{Data2Vec-A} & 313             & \textbf{0.6587}  & 0.5926  & 0.6557    \\
\textbf{DeCoAR 2}   & 90              & \textbf{0.6462}  & 0.5830  & 0.6433    \\
\textbf{W2V2 R}     & 317             & \textbf{0.6470}  & 0.5598  & 0.6132    \\
\textbf{W2V}        & 33              & \textbf{0.6118}  & 0.5385  & 0.6045    \\
\textbf{APC}        & 4               & \textbf{0.6079}  & 0.5433  & 0.6040    \\
\textbf{VQ-APC}     & 5               & \textbf{0.6030}  & 0.5380  & 0.6021    \\
\textbf{TERA}       & 21              & 0.5964  & 0.5416  & \textbf{0.6043}    \\
\textbf{Mockingjay} & 85              & 0.5704  & 0.5344  & \textbf{0.5783}    \\
\textbf{NPC}        & 19              & 0.5701  & 0.5267  & \textbf{0.5822}    \\
\textbf{M CPC}      & 2               & \textbf{0.5272}  & 0.4764  & 0.5262    \\ \midrule
\textbf{FBANK}      & 0               & 0.1442  & \textbf{0.1580}  & 0.1528    \\ \bottomrule
\end{tabular}
\label{tab:cremad_modality_sslms}
\end{table}

\begin{table*}[!t]
\fontsize{7}{9}\selectfont
\vspace{-3mm}
\centering
\caption{Overview of SER performances based on the \textbf{WavLM Large} in the cross-testing conditions. The \textbf{AV} represents ``Audio-Visual''. For column, \textbf{Overall}, we also indicate the lower and upper bound of the confidence interval between 2.75\% and 97.5\% for each result (lower bound, upper bound) using \cite{ferrer_2024} in macro-F1 scores. For other columns, we use sample-based F1-scores.} 
\vspace{-3mm}
\begin{tabular}{@{}c|c|c|cccccc@{}}
\toprule
Train Set & Test Set                  & Overall (lower bound, upper bound)                           & Angry           & Sad             & Disgust         & Fear            & Neutral         & Happy           \\ \midrule
Voice & \multirow{4}{*}{Voice} & \textbf{0.7117   (0.7043,0.7185)} & \textbf{0.7738} & 0.6285          & 0.6301          & \textbf{0.6665} & \textbf{0.9206} & \textbf{0.6508} \\
Face  &                        & 0.6146 (0.6079,0.6215)            & 0.6979          & 0.5583          & 0.5804          & 0.5586          & 0.7488          & 0.5436          \\
AV    &                        & 0.6486 (0.6417,0.6548)            & 0.7394          & 0.6154          & 0.6115          & 0.6289          & 0.7173          & 0.5792          \\ 
All   &                        & 0.6873 (0.6803,0.6938)            & 0.7419          & \textbf{0.6428} & \textbf{0.6394} & 0.6242          & 0.8694          & 0.6059          \\ \midrule
Voice & \multirow{4}{*}{Face}  & 0.5634 (0.5555,0.5703)            & 0.6139          & 0.5075          & 0.5233          & 0.4804          & 0.6713          & 0.5841          \\
Face  &                        & 0.6366 (0.6303,0.6433)            & \textbf{0.6597} & 0.5458          & 0.6061          & \textbf{0.5835} & 0.7414          & 0.6834          \\
AV    &                        & 0.6382 (0.6320,0.6445)             & 0.6559          & 0.5337          & 0.6052          & 0.5775          & \textbf{0.7415} & 0.7154          \\
All   &                        & \textbf{0.6406   (0.6340,0.6471)}  & 0.6548          & \textbf{0.5572} & \textbf{0.6152} & 0.5662          & 0.7201          & \textbf{0.7301} \\ \midrule
Voice & \multirow{4}{*}{AV}    & 0.6418 (0.6350,0.6483)             & 0.7164          & 0.6155          & 0.6100          & 0.5802          & 0.6949          & 0.6337          \\
Face  &                        & 0.6750 (0.6691,0.6812)            & 0.7484          & 0.5593          & 0.6507          & 0.6264          & 0.7628          & 0.7021          \\
AV    &                        & 0.7076 (0.7016,0.7138)            & \textbf{0.7650} & 0.6246          & 0.6705          & \textbf{0.6679} & \textbf{0.7742} & 0.7436          \\
All   &                        & \textbf{0.7085   (0.7025,0.7142)} & 0.7539          & \textbf{0.6398} & \textbf{0.6936} & 0.6442          & 0.7534          & \textbf{0.7662} \\ \bottomrule
% Voice & \multirow{4}{*}{All}   & 0.6882 (0.6814,0.6950)             & 0.7457          & 0.6226          & 0.6515          & 0.6323          & 0.8413          & 0.6361          \\
% Face  &                        & 0.6928 (0.6865,0.6989)            & 0.7711          & 0.5803          & 0.6550          & 0.6572          & 0.7925          & 0.7009          \\
% AV    &                        & 0.7207 (0.7147,0.7269)            & \textbf{0.7864} & 0.6378          & 0.6854          & \textbf{0.6915} & 0.7815          & 0.7418          \\
% All   &                        & \textbf{0.7454   (0.7391,0.7519)} & 0.7829          & \textbf{0.6543} & \textbf{0.7161} & 0.6866          & \textbf{0.8671} & \textbf{0.7418} \\  \bottomrule
\end{tabular}
\label{tab:cross_results}
\vspace{-5mm}
\end{table*}

\textbf{What is the effect of the stimulus modality on the performance of the SER systems?} The results in Table \ref{tab:cremad_modality_sslms} reveal intriguing differences in the performance of the SER model when trained on labels elicited from different modalities.
For the evaluation annotations, \textbf{Voice} utilizes voice-only annotations, \textbf{Face} uses face-only annotations, and \textbf{AV} uses audio-visual annotations. The crucial insight in Table~\ref{tab:cremad_modality_sslms} is the performance variation when different modalities are elicited.
The best overall macro-F1 score in 9 out of 15 experiments was achieved by models trained with the \textbf{Voice}, and 5 out of 15 experiments was achieved by models trained with the \textbf{AV}. These findings highlight the significant impact that the chosen annotation modality can have on the capability of SER systems to recognize emotions from speech accurately. Consequently, when developing SER models, it is crucial to carefully consider the modality used for annotating emotional labels, as this factor can substantially influence the model's ability to capture the nuances of emotions in speech.

\begin{figure}[!b]
\centering
\vspace{-3mm}
\includegraphics[width=0.86\linewidth]{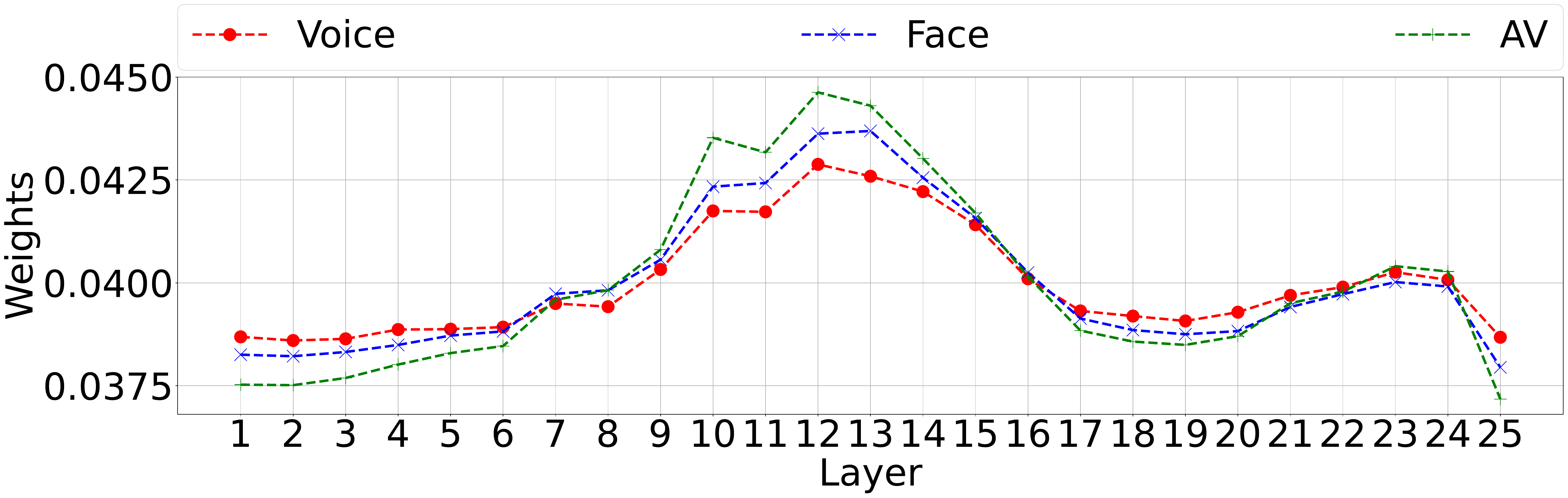}
\vspace{-3mm}
\caption{The layerwise weights of WavLM-based SER systems trained with the labels elicited by various modalities. The \textbf{Face}, \textbf{Voice}, and \textbf{AV} represent face-only, voice-only, audio-visual, and all-inclusive, respectively.} 
\label{fig:layerwise}
\end{figure}

\textbf{Is there a difference between SER systems trained with the labels elicited by different conditions based on the layerwise weights analysis?} We conduct a layerwise analysis to understand the importance of different layers in the \textbf{WavLM Large}-based SER models trained with the labels elicited by various modalities. We extract the layer weights from the best checkpoint of each model and normalize them using the softmax function to ensure values between 0 and 1. We average the layerwise weights across multiple partitions of the CREMA-D dataset. Fig.~\ref{fig:layerwise} plots the layer weights across all models trained with emotion labels elicited by various modalities (voice-only, face-only, and audio-visual). The models tend to assign higher weights to the 10th to 15th layers, suggesting that these middle layers encode more emotional information than the earlier or later layers. Interestingly, the model trained with voice-only labels exhibits more balanced weights across layers than those taught with labels elicited by other modalities.

\textbf{Which labels elicited by various modalities are the most effective for SER systems?} 
We choose the best model (\textbf{WavLM Large}) in Table~\ref{tab:cremad_modality_sslms} as the backbone model for the experiments. 
Table~\ref{tab:cross_results} summarizes the performance of SER systems trained with labels elicited by various modalities and evaluated on test sets defined by labels elicited by different modalities. The ``Train Set'' column shows the models trained by which label type (\textbf{Voice}, \textbf{Face}, \textbf{AV}, or \textbf{All}), and the ``Test Set'' column shows the testing label type that decides the ground truth of the test sets. The testing conditions are of three types: \textbf{Voice}, \textbf{Face}, and \textbf{AV}. In the \textbf{Overall} column of Table~\ref{tab:cross_results}, we report the macro-F1 scores along with the lower and upper bounds of the confidence interval between 2.75\% and 97.5\%  for each result.
%, calculated using the toolkit proposed by \cite{ferrer_2024}. 
Additionally, we present the sample-based F1 scores for the recognition performance of each emotion.

Interestingly, when tested on the voice-only condition (\textbf{Voice}), the SER system trained with voice-only labels achieved promising results compared to models trained with other modalities. This finding suggests that voice-only labels are suitable for training SER systems, as the input is speech-only. This finding connects to recent benchmark databases (such as MSP-PODCAST \cite{Lotfian_2019_3}) that use audio-only stimuli for labels. Furthermore, we observed that the model trained with the proposed \emph{all-inclusive} label set (\textbf{All}) performed better on sad and disgusted emotions than the one trained with voice-only labels (\textbf{Voice}).  

Regarding the testing conditions using face-only (\textbf{Face}) and audio-visual (\textbf{AV}), our proposed label type (\textbf{All}) results in the best performance, demonstrating the effectiveness of the proposed label sets. The models trained with the proposed label set perform best for sad, disgust, and happy emotions on the three testing conditions.

\section{Discussion and Limitations}
While the study by Paulmann et al. \cite{Paulmann_2011} suggests that humans have enhanced emotion recognition with multimodal stimuli compared to single modalities, our research found that audio-only labels are the most effective for training SER systems when only speech input
is available. Systems were evaluated using audio-only, visual-only,
audio-visual, and all-inclusive labels, with the audio-only approach
proving to be the most optimal. Multimodal emotion systems are
not used since current SER systems predominantly rely exclusively
on audio input. Moreover, to ensure that emotion predictions closely
resemble human perceptions, speech-only emotion recognition evaluations
must use labels derived from voice-only contexts. Additionally,
the findings from the mentioned study might not apply to other emotion
recognition systems, such as those based on facial expressions,
text, or audio-visual inputs.
Besides, our experiments have one limitation: they were conducted on a single emotion database, as no other publicly available databases provide emotional annotations across various stimuli.

% While the study by Paulmann et al. \cite{Paulmann_2011} suggests that humans recognize emotions better when presented with multi-modal stimuli compared to single modalities, our findings indicate that audio-only labels are the most effective for training SER systems among the modalities we explored (audio-only, visual-only, audio-visual, and all-inclusive). The main reason is that current SER systems only take audio as input, whereas humans can decode emotional cues from multiple modalities. Emotion predictions should closely mirror human perceptions, and hence, the evaluation of speech-only emotion recognition should employ labels derived from voice-only contexts. In addition, the paper's findings might not be suitable for other emotion recognition systems, such as facial emotion recognition, text emotion recognition, and audio-visual emotion systems. 
%Besides, our experiments have one limitation: they were conducted on a single emotion database, as no other publicly available databases provide emotional annotations across various stimuli.

\section{Conclusion and Future Work}

This work compares SER systems (based on 14 SSLMs) trained with labels elicited from various emotional stimuli (multi-modal and uni-modal). The results show that the different modalities of emotional stimuli can significantly impact the performance of SER systems. Also, we propose an \emph{all-inclusive label} set that combines labels elicited by multi-modal and uni-modal emotional stimuli. The SER systems trained on the proposed \emph{all-inclusive label} set achieved the best performance on test sets for facial-only and audio-visual scenarios. Moreover, the SER systems trained solely on labels elicited by the voice-only stimuli provided promising results on the voice-only test condition, suggesting that voice-only training is preferable for speech-only applications. The findings connect to recent benchmark databases (such as MSP-PODCAST) that use audio-only stimuli for labels and align with how humans perceive emotions through voice alone. Also, the findings suggest that speech-only SER systems find it challenging to interpret emotional ratings derived from audio-visual or face-only modalities, as these lack the inherent emotional signals in voice. In future work, we plan to incorporate systems that can take different modalities, such as audio and video inputs. %, since humans can decode emotions from multiple modalities.

\section*{Acknowledgment}
\label{sec:ack}
This work was supported by the NSTC under Grant 113-2634-F-002-003. We thank Professor Hung-yi Lee for his valuable comments.

% We thank the anonymous reviewers and Professor Carlos Busso for their valuable comments.  
%We also thank the National Center for High-performance Computing (NCHC) for providing computational and storage resources.

% \small
\bibliographystyle{IEEEtran}
\bibliography{IEEEabrv,refs}

\end{document}